# The rotating Morse potential model for diatomic molecules in the J-matrix representation:
## II. The S-matrix approach

I. Nasser, M. S. Abdelmonem and H. Bahlouli
*Physics Department, King Fahd University of Petroleum & Minerals, Dhahran 31261, Saudi Arabia*

A. D. Alhaidari
*Shura Council, Riyadh 11212, Saudi Arabia*

This is the second article in which we study the rotating Morse potential model for diatomic molecules using the tridiagonal J-matrix approach. Here, we improve further the accuracy of computing the bound states and resonance energies for this potential model from the poles of the S-matrix for arbitrary angular momentum. The calculation is performed using an infinite square integrable basis that supports a tridiagonal matrix representation for the reference Hamiltonian, which is included in the computations analytically without truncation. Our method has been applied to both the regular and inverted Morse potential with favorable results in comparison with available numerical data. We have also shown that the present method adds few significant digits to the accuracy obtained from the finite dimensional approach (e.g. the complex rotation method). Moreover, it allows us to handle easily both analytic and non-analytic potentials as well as 1/r singular potentials.

It is well known that the Morse potential model describes very well the vibrations of diatomic molecules [1]. This is because such systems can be modeled by two positive charges (the two atomic nuclei) that produce the Coulomb potential in which the electronic cloud moves. A similar type of modeling was also applied to nuclear molecules [2]. However, in this case there are no well defined centers and the potential in which the nucleons move arises from the sum of interaction between all nucleons. In the language of energy scales, the electronic excitation energy, the vibration energy and the rotational molecular energy for diatomic molecules are widely separated energy scales, hence making the three degrees of freedom completely uncoupled. For a nuclear molecule, on the other hand, the single nucleon excitation and rotational energies are comparable for even the lowest angular momentum. Thus, it is expected that the energy associated with vibrations of the inter-nuclear coordinates will also be of the same order of magnitude. This will lead to a more complex energy spectrum reflecting the much stronger coupling between the intrinsic, vibrational and rotational degrees of freedom of nuclear molecules.

The Morse potential is an element in the class of potentials for which an analytic solution of the Schrödinger equation exists for zero angular momentum [3]. It is written as

$$V_M(r) = V_0 \left[ e^{-2\alpha(x-1)} - 2 e^{-\alpha(x-1)} \right]; \quad x = r/r_0 \tag{1}$$

where $V_0$ is the strength of the potential (or the dissociation energy in the context of diatomic molecules), $r_0$ is the equilibrium intermolecular distance and $α$ is a positive number controlling the decay length of the potential. For $V_0 > 0$, this potential has a minimum of $-V_0$ at $r = r_0$ and it is called the regular Morse potential. For $V_0 < 0$, it has a maximum of $-V_0$ there and it is called the inverted Morse potential. Asymptotically (as *r* becomes very large) it goes to zero. The rotating Morse potential is the sum of the repulsive centrifugal potential barrier, $\frac{\hbar^2}{2m}\frac{\ell(\ell+1)}{r^2}$, and the Morse potentials (1). Generally, this effective potential shows a valley followed by a potential barrier which will then have two major effects. For the inverted Morse potentials where $V_0 < 0$ some resonances will be developed and the number of bound states will be reduced as $\ell$ increases. There will also be a critical angular momentum value, $\ell_c$, beyond which no bound states are found and this critical value will depend on the potential parameters $V_0$ and *α*. Many numerical approaches using various approximation techniques have been proposed for solving the rotating Morse potential. These were extensively implemented with varying degrees of accuracy and stability [4]. Other semi-analytic methods have also been used such as the Nikiforov-Uvarov method [5] and the asymptotic iteration methods [6].

In our previous work [7] we have used the tridiagonal representation approach inspired by the J-matrix method [8] to compute the bound state energy spectrum of four different types of diatomic molecules: $H_2$, LiH, HCl and CO. Our approach provided an alternative method for obtaining the bound states energies for these diatomic molecules with improved accuracy and could easily be extended to other molecules. The numerical results have been compared favorably with those obtained using other approximation schemes [4]. In this work, we extend the approach to handle not only the bound states but also resonances associated with the rotating Morse potential model.

The direct way to compute the resonances is based on the accurate definition of the resonances as being the poles of the scattering S-matrix in the complex energy plane. One can show that each element of the S-matrix is singular at the complex resonance energy *E*
$$S^{-1}(E) = 0; \quad E = E_R \pm i\, E_I. \tag{2}$$
This condition is sufficient for obtaining the resonance position $E_R$ and width $\Gamma = 2E_I$. Several methods to find the complex resonance energies of a given scattering Hamiltonian are available. There are many techniques that enable us to evaluate the S-matrix. One such approach is to use the Jost function and its analytic properties [9]. In another approach, Yamani and Abdelmonem [10] showed how to calculate the S-matrix at the real Harris energy eigenvalues and, subsequently perform analytic continuation of the S-matrix to the complex energy plane. The required resonance information is then extracted from the analytically continued S-matrix. In our present work, we will evaluate the resonances and bound state energies associated with the rotating Morse potential by combining the properties of the S-matrix, the complex rotation method [11] and the analytical and computational power of the J-matrix approach [8]. In our previous work [7], we have restricted our calculation to a finite dimensional subspace spanned by a subset of the J-matrix square integrable basis. Due to the finiteness of the dimensional space, the accuracy of our numerical results is somewhat limited. Any additional improvement requires larger spaces, consequently more computational time and less stability. Since the objective is to increase the accuracy and improve the efficiency in locating the resonance positions and widths without extending too much the computing time or reducing numerical stability, we have opted in the present work to use the analytic power of the J-matrix approach. It will enable us to include, without truncation, all of the exactly solvable part of the Hamiltonian as an infinite tridiagonal matrix tail (usually referred to as the $H_0$ problem in the J-matrix literature). For the present problem, $H_0$ is just the kinetic energy operator (i.e., the sum of the second order radial differential operator and the orbital term).

Direct study of resonances is usually done in the complex energy plane. Resonance energies are the subset of the poles of the S-matrix which are located in the lower half of the complex energy plane. One way to uncover these resonances, which are "hidden" below the real line in the complex $E$-plane, is to use the complex scaling (complex rotation) method [11]. This method exposes the resonance poles and makes their study and manipulation easier. The subset of eigenvalues that corresponds to resonance and bound states spectra remain stable against variations in all computational parameters.

The time-independent Schrödinger wave equation for a point particle in a spherically symmetric potential $V(r)$ reads as follows

$$(H-E)\chi = \left[ -\frac{\hbar^2}{2m}\frac{d^2}{dr^2} + \frac{\ell(\ell+1)\hbar^2}{2mr^2} + V(r) - E \right]\chi = 0, \qquad (3)$$

where $\chi(r)$ is the wavefunction which is parameterized by the potential parameters, $\ell$ and $E$. We expand $\chi$ in an $L^2$ complete basis set $\{\phi_n\}$ which is chosen to make the matrix representation of the reference Hamiltonian, $H_0 \equiv H - V$, tridiagonal. The basis is parameterized by a positive length scale parameter $\lambda$ as $\{\phi_n(\lambda r)\}$, which also allows for more computational freedom. The following choice of basis functions [12] is compatible with the domain of the Hamiltonian, satisfies the desired boundary conditions, and results in a tridiagonal matrix representation for $H_0$

$$\phi_n(r) = a_n (\lambda r)^{\ell+1} e^{-x/2} L_n^\nu(x); \qquad n = 0,1,2,.. \qquad (4)$$

where $x = \lambda r$ and $\nu = 2\ell + 1$ in the "Laguerre basis", whereas $x = (\lambda r)^2$ and $\nu = \ell + \frac{1}{2}$ in the "oscillator basis". $L_n^\nu(x)$ is the Laguerre polynomial of degree $n$ and $a_n$ is the normalization constant $\sqrt{\Gamma(n+1)/\Gamma(n+\nu+1)}$. The reference Hamiltonian $H_0$ in this representation, which is at the heart of the J-matrix approach, is accounted for in full. On the other hand, the short-range Morse potential $V$ is approximated by its matrix elements in an adequate subset of the basis using the Gauss quadrature approach [13]. That is, the matrix representation of the full Hamiltonian becomes

$$H_{nm} \cong \begin{cases} (H_0)_{nm} + V_{nm} & ; \quad n,m \leq N-1 \\ (H_0)_{nm} & ; \quad n,m > N-1 \end{cases}. \qquad (5)$$

Such a representation is the fundamental underlying feature of the J-matrix [8] method. As it is obvious from (5), the reference Hamiltonian, which includes the orbital term, is not truncated as is usually done in the finite calculation methods. This full account of the reference Hamiltonian should result in a substantial improvement on the accuracy of the results especially for large angular momenta. This constitutes the real power of the S-matrix approach that we are proposing. Moreover, in finding resonances and bound state energies we use the "direct method" based on the J-matrix calculation of the scattering matrix in the complex energy plane. Bound states are associated with negative real poles of the S-matrix while resonances are associated with complex poles that have positive real parts and negative imaginary parts. The S-matrix, in the J-matrix approach, is defined by [10]

$$S(E) = T_{N-1}(E) \frac{1 + g_{N-1,N-1}(E) J_{N-1,N}(E) R_N^-(E)}{1 + g_{N-1,N-1}(E) J_{N-1,N}(E) R_N^+(E)}, \text{ where} \quad (6a)$$

$$g_{N-1,N-1}(z) = D_v^N \sum_{n=0}^{N-1} \frac{\Lambda_{N-1,n}^2}{\varepsilon_n - z} = D_v^N \left[ \prod_{m=0}^{N-2} (\tilde{\varepsilon}_m - z) \Big/ \prod_{n=0}^{N-1} (\varepsilon_n - z) \right], \text{ and} \quad (6b)$$

$$T_n = \frac{c_n - i s_n}{c_n + i s_n} ; \quad R_n^{\pm} = \frac{c_n \pm i s_n}{c_{n-1} \pm i s_{n-1}}. \quad (6c)$$

$g_{nm}(z)$ is the inverse of the matrix $(H - z)$ where $H$ is the $N \times N$ finite matrix given by the first line in (5). The eigenvalues and the corresponding normalized eigenvectors of the finite matrix $H$ are denoted by $\{\varepsilon_n\}_{n=0}^{N-1}$ and $\{\Lambda_{nm}\}_{n,m=0}^{N-1}$, respectively. $\{\tilde{\varepsilon}_n\}_{n=0}^{N-2}$ are the eigenvalues of the truncated $H$ obtained by removing the last row and last column. In the two bases (the Laguerre and oscillator), the matrix representation of the reference wave operator, whose elements are defined by $J_{m,n} = \langle \phi_m | (H_0 - E) | \phi_n \rangle$, is tridiagonal and symmetric. The quantities $s_n$ and $c_n$ are the expansion coefficients of the two independent asymptotic solutions of the reference wave equation $(H_0 - E)\psi = 0$. In this J-matrix terminology, the two independent asymptotic solutions of the reference problem are written as

$$|S\rangle = \sum_{n=0}^{\infty} s_n |\phi_n\rangle; \quad |C\rangle = \sum_{n=0}^{\infty} c_n |\phi_n\rangle, \quad (7)$$

and are usually called the "sine-like" and "cosine-like" solutions, respectively. Table 1 gives all the elements needed to calculate the S-matrix (6a). Using the recursion relation, we can obtain $T_{N-1}(E)$ and $R_N^{\pm}(E)$ from $T_0(E)$ and $R_1^{\pm}(E)$ recursively in the form of a continued fraction as shown in [10].

In the following we give the recipe of our procedure for calculating the bound and resonance energies for the rotating Morse potential for different values of potential parameters and $\ell$. Our calculation strategy is based on the following procedure: First we use the complex rotation method [11] to calculate these energies for the given physical parameters. This step is very important in our approach since these values constitute the seed for the S-matrix computations. Now we proceed to find the roots of $S^{-1}(E)$ matrix. For this purpose, the built-in iteration procedure in the computational software package MATHEMATICA (version 5) was used [14]. Due to this iteration, the accuracy of the S-matrix results depends crucially on the accuracy of the seeds that were obtained using the complex rotation approach. In the results of our calculation, the number of significant digits shown is usually obtained for the optimum values of the numerical parameters $\lambda$ and $\theta$ and for a given choice of the dimension parameter $N$.

As a continuation of our efforts deployed in paper I [7] we apply the strategy outlined above to the study of both resonances and bound states. However, numerical computations for the diatomic molecules $H_2$, LiH, HCl and CO with the associated model parameters given in [7] produced only bound states. Therefore, we need to choose appropriate potential parameters that can give rise to resonances. As explained in the introduction the inverted Morse potential gives rise to resonances and bound states even for $\ell = 0$ and it is for this reason that it was considered in the literature [15].

In Table 2a, we show the bound and resonance energies for the inverted Morse potential obtained using the J-matrix for the same parameters and units used in [15]. That is, in the potential model (1), we took $V_0 = -6$ fm$^{-2}$, $\alpha = 0.3$ fm$^{-1}$, $r_0 = 4.0$ fm, and $\ell = 0$. The energies in the table are given in fm$^{-2}$ as in [15]. One bound state and 15 resonances were obtained for this potential using the S-matrix approach in the Laguerre basis. Our results complement those in [15] where only one bound state and two resonances were given (it is worth mentioning that the resonant at $E_R = 1.1783$ in [15] is not consistent with the $k$-value given in the same reference). Nine of the 15 resonances are located in the fourth quadrant of the complex energy plane while six are in the third quadrant. Except for the first two resonances, which are shallow, all others resonances are deep. It is to be noted that for deep resonances, we need larger values of $N$ and suitable values of $\lambda$ and $\theta$ to reach the desired accuracy. It is also worth mentioning that negative energy resonances have recently found practical application in ultra-cold atomic collisions at negative energies [16]. Similar negative energy resonances were discussed in the past by one of the Authors [17]. In Table 2b, we show the results of our calculation using the same parameters as in Table 2a but with nonzero angular momentum ($\ell = 1$). Additionally, in Table 3, we use the oscillator basis and choose a different set of model parameters (in atomic units) and for $\ell = 2$.

The numerical results for the resonances and bound states shown in Tables 2 and 3, which are generated by the S-matrix approach, are very close to those obtained by complex rotation. Hence the effect of the infinite tail resulting from the J-matrix approach for this model has no substantial contribution. On the other hand, we expect that the infinite tail originating from the $H_0$ contribution will be more important for higher values of the angular momentum and/or shorter potential range (large $\alpha$). To study quantitatively the tail effect we use the following numerical approximation for the Hamiltonian

$$H_{nm} \cong \begin{cases} (H_0)_{nm} + V_{nm} & ; \quad n, m \leq N - 1 \\ (H_0)_{nm} & ; \quad M + N - 1 \geq n, m > N - 1 \end{cases} \qquad (8)$$

which restricts the potential contribution to an $N \times N$ subspace while the $H_0$ part of the Hamiltonian is extended to $(N+M) \times (N+M)$ space to allow for an extra $M$-dimensional tridiagonal tail contribution from $H_0$. In the S-matrix approach, this tail goes to infinity but we will see numerically that the tail contribution in our model reaches its asymptotic contribution very quickly. Specifically, the asymptotic limit is already reached for $M = 10$. Figure 1 shows the results of resonance calculations with the parameters (in atomic units $\hbar = m = 1$): $V_0 = -10$, $r_0 = 1$, $\alpha = 0.5$ and $\ell = 5$ using the Laguerre basis with $\lambda = 15$ and $\theta = 0.9$ radians. First, we computed the resonances associated with this potential for a large value of $N$, ($N = 100$ and $M = 0$) shown as crosses in Fig. 1 where they serve as a reference for comparison to the values generated with smaller values of $N$ and increasing length of the tail. We notice from the figure that the tail effect is more pronounced for deep resonances. A similar analysis can be done by looking at the numerical values in a tabular form for these resonances in the presence of a tail, which shows that the tail effect is really confined to improving the last couple of digits.

In summary, we extended our approach in [7] to the computation of the bound states and resonance energies, as being the poles of the S-matrix, using the power of the J-matrix technique that includes a partial contribution of the potential but full analytic contribution of the reference Hamiltonian. Our approach could easily be generalized to handle other short-range and even 1/r singular potentials such as the Yukawa and Hulthén potentials, to mention only a few [18]. These results suggest that our present approach is of comparable accuracy to the complex rotation approach. However, the J-matrix approach can easily handle non-analytic and 1/r singular potentials, which cannot be handled with great accuracy using other approaches and cannot be handled easily using the complex rotation method. We have also studied the importance of the tail contribution inherent in the J-matrix approach and showed that the tail effect will improve the accuracy of the resonances by few significant digits for the present model.

In the future third article of this series, we will investigate a generalized three-parameter Morse potential model that should be more suitable and give a higher degree of freedom for the description of various diatomic molecules. This three-parameter Morse potential reads as follows

$$V_{GM}(r) = V_0 \left[ e^{-2\alpha(r-r_0)/r_0} - 2\beta e^{-\alpha(r-r_0)/r_0} \right], \qquad (9)$$

where $\beta$ is a new dimensionless parameter whose value is unity for the regular Morse potential. As shown in paper I, this generalized potential results in an exact S-wave ($\ell = 0$) solution for the time-independent Schrödinger equation.

**Acknowledgment**
We acknowledge the support provided by the Physics Department at King Fahd University of Petroleum & Minerals under project FT-2008-10. We are also grateful to "Khaled Technical & Commercial Services" (KTeCS) for the generous support. We also appreciate the comments of the Referees that resulted in significant improvements on the presentation of the work.

**References**

[1]  P. M. Morse, Phys. Rev. **34,** 57 (1929); S. H. Dong, R. Lemus, A. Frank, Int. J. Quant. Phys. **86,** 433 (2002) and references therein.
[2]  R. R. Betts and A. H. Wuosmaa, Rep. Prog. Phys. **60,** 819 ( 1997).
[3]  S. Flügge, *Practical Quantum Mechanics*, vol. I (Springer-Verlag, Berlin, 1994).
[4]  C. L. Pekeris, Phys. Rev. **45**, 98 (1934); R. Herman, R. J. Rubin, Astrophys. J. **121**, 533 (1955); M. Duff, H. Rabitz, Chem. Phys. Lett. **53**, 152 (1978); J. R. Elsum, G. Gordon, J. Chem. Phys. **76**, 5452 (1982); E. D. Filho, R. M. Ricotta, Phys. Lett. A **269**, 269 (2000); F. Cooper, A. Khare, and U. Sukhatme, Phys. Rep. **251**, 267 (1995); D. A. Morales, Chem. Phys. Lett. **394**, 68 (2004); J. P. Killingbeck, A. Grosjean, and G. Jolicard, J. Chem. Phys. **116**, 447 (2002); T. Imbo and U. Sukhatme, Phys. Rev. Lett. **54**, 2184 (1985); M. Bag, M. M. Panja, R. Dutt and Y. P. Varshni, Phys. Rev. A **46**, 6059 (1992).
[5]  A. F. Nikiforov and V. B. Uvarov, *Special Functions of Mathematical Physics* (Basel, Birkhausr, 1988); C. Berkdemir and J. Han, Chem. Phys. Lett. **409**, 203 (2005); C. Berkdemir, Nucl. Phys. A **770**, 32 (2006).
[6]  O. Bayrak and I. Boztosum, J. Phys. A **39**, 6955 (2006).
[7]  I. Nasser, M. S. Abdelmonem, H. Bahlouli and A. D. Alhaidari, J. Phys. B **40**, 4245 (2007).

[8]  E. J. Heller and H. A. Yamani, Phys. Rev. A **9**, 1201 (1974); H. A. Yamani and L. Fishman, J. Math. Phys. **16**, 410 (1975); A. D. Alhaidari, E. J. Heller, H. A. Yamani, and M. S. Abdelmonem (eds.), *The J-matrix method: developments and applications* (Springer-Verlag, Dordrecht, 2008).

[9]  S. A. Sofianos and S. A. Rakityansky, J. Phys. A **30**, 3725 (1997); S. A. Rakityansky, S. A. Sofianos and N. Elander, J. Phys. A **40**, 14857 (2007); H. M. Nussenzveig, *Causality and Dispersion Relations* (Academic, London, 1972).

[10] H. A. Yamani and M. S. Abdelmonem, J. Phys. B **30**, 1633 (1997); **30**, 3743 (1997).

[11] J. Aguilar and J. M. Combes, Commun. Math. Phys. **22**, 269 (1971); Y. K. Ho, Phys. Rep. **99**, 1 (1983).

[12] A. D. Alhaidari, Ann. Phys. (NY) **317,** 152 (2005).

[13] See, for example, Appendix A in: A. D. Alhaidari, H. A. Yamani, and M. S. Abdelmonem, Phys. Rev. A **63**, 062708 (2001).

[14] S. Wolfram, *The Mathematica Book*, Wolfram Media, 1999.

[15] G. Rawitscher, C. Merow, M. Nguyen and I. Simbotin, Am. J. Phys **70** , 935 (2002).

[16] S. G. Bhongale, S. J. J. M. F. Kokkelmans and Ivan H. Deutsch, Phys. Rev. A **77**, 052702 (2008).

[17] A. D. Alhaidari, Int. J. Mod. Phys. A **20**, 2657 (2005).

[18] A. D. Alhaidari, H. Bahlouli and M. S. Abdelmonem, J. Phys. A **41**, 032001 (2008).

**Table Captions**

**Table 1:** The explicit form of the kinematic quantities $T_0(E)$, $R_1^\pm(E)$, $J_{N-1,N}(E)$ and $D_\nu^N$ in both Laguerre and oscillator bases. The three-term recursion relations for $s_n$ and $c_n$ (collectively shown as $f_n$) in both bases are also given.

**Table 2a:** Bound and resonance energies calculated using the S-matrix approach in the Laguerre basis for the inverted Morse potential with the parameters: $V_0 = -6$ fm$^{-2}$, $\alpha = 0.3$ fm$^{-1}$, $r_0 = 4.0$ fm, and $\ell = 0$.

**Table 2b:** Reproduction of Table 2a but for $\ell = 1$

**Table 3:** Bound and resonance energies calculated using the S-matrix approach in the oscillator basis for the following Morse potential parameters (in the units $\hbar = m = 1$): $V_0 = -10$, $\alpha = 2.0$, $r_0 = 1.0$, and $\ell = 2$.

**Table 1**

|  | **Laguerre basis** | **Oscillator basis** |
|---|---|---|
| $T_0(E)$ | $e^{2i\theta}\,\dfrac{{}_2F_1(-\ell,1;\ell+2;e^{2i\theta})}{{}_2F_1(-\ell,1;\ell+2;e^{-2i\theta})};\quad \cos\theta = \dfrac{8E-\lambda^2}{8E+\lambda^2}$ | $s_0 = \dfrac{\sqrt{2\pi}}{\Gamma(\nu+1)} e^{-x/2} x^{(2\nu+1)/4};\quad x = 2E/\lambda^2$ <br> $c_0 = \dfrac{\sqrt{2/\pi}}{\nu} e^{-x/2} x^{(1-2\nu)/4}\,{}_1F_1(-\nu,1-\nu;x)$ |
| $R_1^\pm(E)$ | $\dfrac{1}{(\ell+2)} e^{\mp i\theta}\,\dfrac{{}_2F_1(-\ell,2;\ell+3;e^{\mp i\theta})}{{}_2F_1(-\ell,1;\ell+2;e^{\mp i\theta})}$ | $s_1 = -\dfrac{\sqrt{2\pi}}{\Gamma(\nu+2)}(\nu+1-x) e^{-x/2} x^{(2\nu+1)/4}$ <br> $c_1 = -\dfrac{\sqrt{2/\pi}}{\nu(\nu+1)} e^{-x/2} x^{(1-2\nu)/4}\,{}_1F_1(-\nu-1,1-\nu;x)$ |
| $J_{N-1,N}(E)$ | $(E+\lambda^2/8)\sqrt{N(N+\nu)}$ | $(\lambda^2/2)\sqrt{N(N+\nu)}$ |
| $D_\nu^N$ | $N+\nu$ | $1$ |
| Recursion Relation | $2(\cos\theta)(n+\ell+1)f_n - \sqrt{n(n+2\ell+1)}\,f_{n-1}$ <br> $-\sqrt{(n+1)(n+2\ell+2)}\,f_{n+1} = 0;\quad n \geq 1$ | $(2n+\ell+\tfrac{3}{2}-x)f_n - \sqrt{n(n+\ell+\tfrac{1}{2})}\,f_{n-1}$ <br> $-\sqrt{(n+1)(n+\ell+\tfrac{3}{2})}\,f_{n+1} = 0;\quad n \geq 1$ |

**Table 2a**

| $N$ | $\lambda$ | $E = E_R - i E_I$ (fm$^{-2}$) | Rawitscher [15] |
|---|---|---|---|
| 25 | 25 | –8.108988161869 | –8.1090 |
| 30 | 35 | 1.177806385671 –i $2.01 \times 10^{-13}$ | 1.1783 |
| 35 | 40 | 5.625155807690 –i 0.035131892716 | 5.62516 –i 0.0351319 |
| 45 | 40 | 6.891102707478 –i 1.319371415391 | |
| 50 | 40 | 7.318247116408 –i 3.588665807635 | |
| 50 | 45 | 7.111103616788 –i 6.071543045928 | |
| 55 | 50 | 6.362668116127 –i 8.600530465794 | |
| 60 | 50 | 5.144646234924 –i 11.09595615976 | |
| 60 | 50 | 3.512295278617 –i 13.51513945824 | |
| 75 | 65 | 1.509512035105 –i 15.83342896153 | |
| 75 | 65 | –0.827814446009 –i 18.03582315814 | |
| 75 | 70 | –3.469692551465 –i 20.11276308840 | |
| 75 | 70 | –6.39069522808 –i 22.05787612132 | |
| 85 | 80 | –9.56907064438 –i 23.8667476308 | |
| 85 | 80 | –12.986084802 –i 25. 53625088 | |
| 85 | 80 | –16.625492 –i 27.064165 | |

**Table 2b**

| $N$ | $\lambda$ | $E = E_R - i E_I$ (fm$^{-2}$) |
|---|---|---|
| 30 | 30 | –1.812991439373 |
| 40 | 35 | 4.343378629186 –i 0.000086853698 |
| 40 | 35 | 6.665075458079 –i 0.531572234393 |
| 45 | 35 | 7.364686543876 –i 2.568333087462 |
| 45 | 35 | 7.436775874472 –i 4.992463411191 |
| 60 | 45 | 6.935679941774 –i 7.522073704643 |
| 80 | 60 | 5.938612579913 –i 10.04794300288 |
| 80 | 60 | 4.506823029849 –i 12.51388896973 |
| 80 | 60 | 2.688440123933 –i 14.88764214149 |
| 80 | 60 | 0.522010509971 –i 17.14952298728 |
| 80 | 60 | –1.960934522180 –i 19.28710891784 |
| 80 | 60 | –4.734098232106 –i 21.29242302118 |
| 80 | 60 | –7.775172363055 –i 23.16032781447 |
| 80 | 60 | –11.06495109797 –i 24.88754934339 |

**Table 3**

| $N$ | $\lambda$ | $E = E_R - \mathrm{i}\, E_I$ (a.u.) |
|----|----|----|
| 70 | 10 | −30.4138814 |
| 70 | 8  | 10.9260145 −i 0.302741 |
| 70 | 8  | 17.123950 −i 12.502698 |
| 70 | 8  | 11.052057 −i 32.190569 |
| 80 | 9  | −5.037642 −i 52.540664 |

**Figure Caption**

**Figure 1:** Resonance plots identified with the pair (*N*,*M*), where *N* is the size of the potential matrix and *N*+*M* is the size of the $H_0$ matrix. The crosses show the most accurate location of the resonances (corresponding to *N* = 100 and *M* = 0). The grid scale is 10.0 energy units ($V_0$ units).

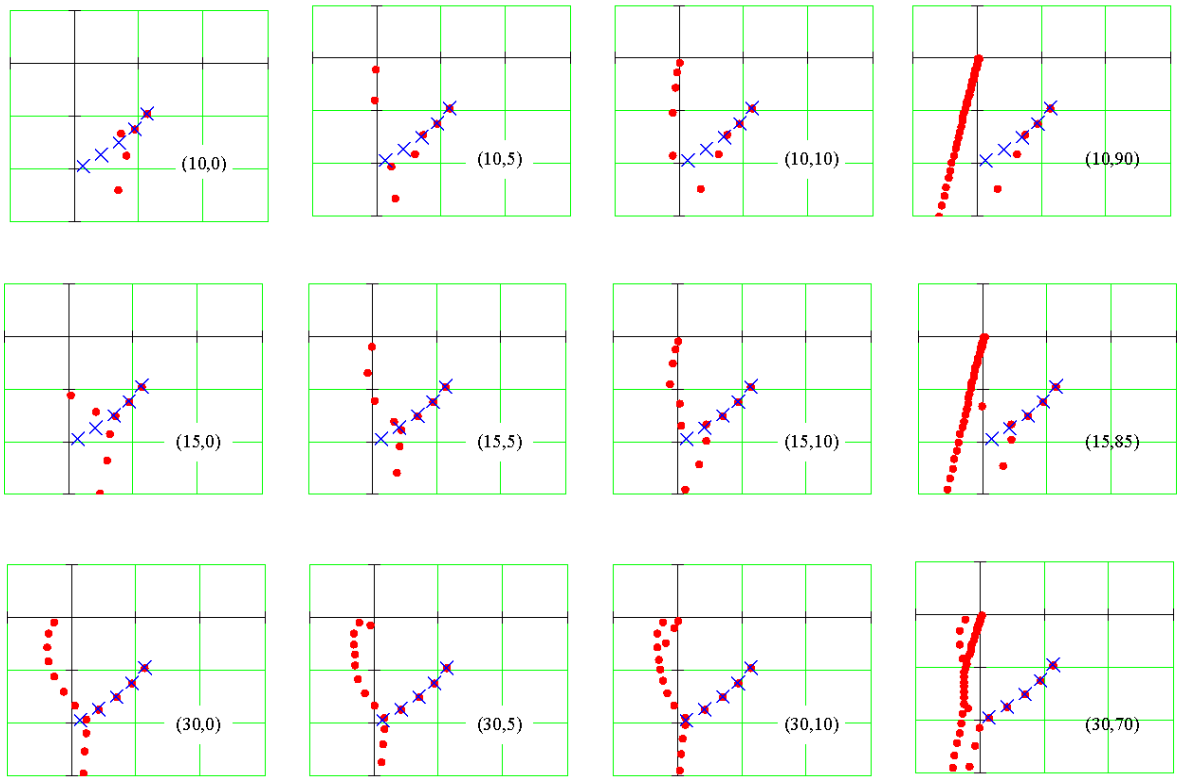

**Fig. 1**